# CUAB: Convolutional Uncertainty Attention Block Enhanced the Chest X-ray Image Analysis

Chi-Shiang Wang, PhD, Fang-Yi Su, MD, Tsung-Lu Michael Lee, PhD, Yi-Shan Tsai, MD, and Jung-Hsien Chiang, PhD, IEEE senior member

*Abstract*—**In recent years, convolutional neural networks (CNNs) have been successfully implemented to various image recognition applications, such as medical image analysis, object detection, and image segmentation. Many studies and applications have been working on improving the performance of CNN algorithms and models. The strategies that aim to improve the performance of CNNs can be grouped into three major approaches: (1) deeper and wider network architecture, (2) automatic architecture search, and (3) convolutional attention block. Unlike approaches (1) and (2), the convolutional attention block approach is more flexible with lower cost. It enhances the CNN performance by extracting more efficient features. However, the existing attention blocks focus on enhancing the significant features, which lose some potential features in the uncertainty information.**

**Inspired by the test time augmentation and test-time dropout approaches, we developed a novel convolutional uncertainty attention block (CUAB) that can leverage the uncertainty information to improve CNN-based models. The proposed module discovers potential information from the uncertain regions on feature maps in computer vision tasks. It is a flexible functional attention block that can be applied to any position in the convolutional block in CNN models. We evaluated the CUAB with notable backbone models, ResNet and ResNeXt, on a medical image segmentation task. The CUAB achieved a dice score of 73% and 84% in pneumonia and pneumothorax segmentation, respectively, thereby outperforming the original model and other notable attention approaches. The results demonstrated that the CUAB can efficiently utilize the uncertainty information to improve the model performance.**

## I. INTRODUCTION

CONVOLUTIONAL neural networks (CNNs) are widely used in various applications, such as object detection and localization [1]–[5] and image segmentation [6]–[8]. At each convolutional layer, the convolutional kernels extract the important features and produce new feature maps that aggregate all channel information from the previous layer. Therefore, methods that can increase the feature map variety and improve the kernels to extract more efficient features have become a popular research topic. Some studies, such as residual networks [9], have attempted to adjust the network architecture to ensure that the convolutional layer extends the previous information, thereby preventing the loss of important features on the next layer. Unlike highway networks [10], residual networks can pass the previous information to the next stage in its original form, which reduces the trainable parameters and prevents the degradation problem using the identity shortcut [11]. In contrast, some studies focused on reusing the features and increasing the feature utilities, such as DenseNet [12]. In DenseNet, a dense block concatenates the feature maps from each convolutional layer; this implies that when the model updates the kernel of the last layer, backpropagation affects the kernel of the first layer simultaneously. It should be noted that the aforementioned studies attempted to enhance the performance of the CNN model by redesigning the model architecture.

However, in recent years, some studies have focused on creating a machine that can make its own decisions to build a more efficient CNN model, named AutoML [13], [14]. The idea of this research field is to leverage the optimizing or searching algorithm to determine a better model structure or forwarding path without any manual design. In AutoML, the machine can decide the size of the kernel, activation function, number of kernels, and the stride of the kernel, such as EfficientNet [15] and AdaNet [16]. Therefore, the design logic of the final model would be significantly different from human thinking. Although the model structure can be optimized by AutoML and obtain better performance, it is difficult to implement it in practical applications owing to the high cost and training device requirements.

Despite model structure can be optimized and obtain good performance, AutoML is more difficult to extend to other models. Therefore, several studies [17]–[20] have re-analyzed the basic convolutional layer to determine how the feature variety can be increased and more efficient features can be extracted based on the existing backbone models; this was named the convolutional attention block module. The concept of the convolutional attention block module is similar to that utilized by [21], wherein the inner product was used to calculate the similarity between the context vector and each hidden state vector to determine which vectors were more important; then, the weighted-sum was applied to each vector to enrich the

C. S. Wang, F. Y. Su and J. H. Chiang are with Department of Computer Science and Information Engineering in National Cheng Kung University, No.1, University Road, Tainan City 701, Taiwan (R.O.C). (email: {hyaline0317, fangyi}@iir.csie.ncku.edu.tw, jchiang@mail.ncku.edu.tw)

T. L. Michael Lee is with Department of Information Engineering in Kun Shan University, No.195, Kunda Rd., Yongkang Dist., Tainan City 710303, Taiwan (R.O.C.) (email: t097000085@g.ksu.edu.tw)

Y. S. Tsai is with Department of Medical Imaging in National Cheng Kung University Hospital, No.1, University Road, Tainan City 701, Taiwan (R.O.C). (email: n506356@gmail.com)
.

context vector. There are several advantages to this type of approach. One of the advantages is that it is a flexible method that can be extended to different models with fewer or no additional trainable parameters. Different from the attention mechanism in recurrent neural networks (RNNs) [21], the convolutional attention block must consider two different types of information on the feature map, i.e., channel-wise and spatial-wise information. In channel-wise attention, it focuses on the channel feature maps that can provide more information. In contrast, the spatial-wise information represents the spatial information between the components and feature maps. To improve the spatial information, more relative information should be captured for specific components; this is called spatial-wise attention. One of the most famous approaches is the "Squeeze-and-Excitation Network" [19], which intelligently fuses the channel-wise and spatial-wise information in an attention block; this is further described in the related work.

Nevertheless, the existing approaches are more concentrated on significant information, which will lose the potential information wherein the features are uncertain. In this study, inspired by test-time augmentation (TTA) [22] and test-time dropout (TTD) [23], we found that the uncertainty information is an important factor that can help in improving the training and inference. We can easily apply TTA or TTD in the inference step to enhance the model performance. TTA and TTD utilize the concept of Monte Carlo sampling [24] to obtain several results from one input data. We collected all such results and adopted the ensemble to obtain the final result. Thus, following the concept of TTA and TTD, we designed a novel convolutional attention block module, called the convolutional uncertainty attention block (CUAB). It can make the convolutional layer leverage the uncertainty information during training and inference. The CUAB module is a lightweight and flexible attention block that can easily be added to the existing convolutional block. This approach provides more information to the convolutional kernels and increases the chance of capturing the potential information on feature maps.

This study aimed to design a convolutional attention block that can create convolutional layers to capture the potential information (uncertain information) and apply it to an existing backbone network. The goals of this study include: (1) designing a convolutional attention module for capturing the potential information from the uncertainty information in feature maps; (2) applying it to the existing backbone models, such as ResNet and ResNeXt; and (3) evaluating the model performance with the CUAB module and comparing it with other convolutional attention block modules on two medical image segmentation tasks.

## II. RELATED WORK

In this study, we leveraged the uncertainty information to design a novel convolutional attention block module, CUAB. The concept of CUAB was inspired by the research mentioned in Section 1. Thus, we briefly retrospect the concept of three different ways that can improve the CNNs to extract more efficient features and increase the variety of feature maps. We will briefly discuss the deeper network architecture, automatic architecture search by AutoML, and other notable convolutional attention block modules.

### A. Deeper Network Architecture

The studies by [9], [12], [25] showed that a deeper model can obtain better quality and performance. Therefore, many researchers attempted to make a more efficient convolutional block by increasing the receptive field or feature reusability. Then, the block is repeated to increase the depth of a model to extract more representative features. The authors of [26] proposed Inception V4, which redesigned the convolutional block and pooling method in Inception V3 [27]. They defined different structures of convolutional blocks in different resolution feature maps and received better performance than Inception V3. Moreover, the authors leveraged the concept of ResNet [9] and integrated the residual path with a simpler inception block to make Inception-Resnet-v1 and v2. Both models achieved better performance than Inception V3 and V4. The authors of [28] expanded the concept of neurons in the network to propose an aggregated residual transformation, ResNeXt, which contained three operations: splitting, transformation, and aggregation. The basic concept of the aggerated transformation is to split the input features into several different transformation blocks to perform non-linear transformation. Subsequently, all feature maps from the transformations are aggregated to enhance the feature variety and representativeness. ResNeXt not only increased the performance but also reduced the training parameters.

### B. Automatic Architecture Search

The authors of [29] proposed the idea of creating a machine to search the CNN model architecture by reinforcement learning, called neural architecture search. This is because most of the CNN models are constructed by stacking several convolutional blocks. Therefore, they utilized an RNN to generate a sequence of hyperparameters in the convolutional block, such as the number of kernels, kernel size, and kernel stride size. Although they achieved better performance than the existing model in the Cifar-10 dataset [30], it is difficult to use this concept in a larger dataset because the network architecture search requires a long time and 500 GPU units. Thus, [31] changed the strategy to search for a better basic convolutional block architecture as the target instead of the entire network structure. The entire network architecture still requires humans to decide the depth of the network. However, this method applied NAS to ImageNet and achieved state-of-the-art results. Various studies are still performing investigations in this domain [15], [32], [33]; however, the device requirements and high cost have become limitations in this field.

### C. Convolutional Attention Block Module

A convolutional attention block is a subnetwork that can be embedded in another convolutional block, such as a residual convolutional block [9] and dense convolutional block [12].

The authors of (Hu et al. (2018) designed a famous squeeze-and-excitation network to enhance the ResNeXt [28] performance on classification. The main concept of squeeze-and-excitation is to squeeze the feature maps on the channel-wise information by global average pooling, followed by using a multi-layer perceptron (MLP) [34] to perform non-linear transformation to represent the features. Next, it uses the sigmoid function to activate the squeezed feature as the attention weights and multiplies it with the original feature map to improve feature map representation. The authors of [18] proposed a novel convolutional block that designed two attention blocks, i.e., channel-wise and spatial-wise attention. Although the experimental results showed that their approach improved the performance, it required more additional trainable parameters than the squeeze-and-excitation network. Moreover, Bello et al. (2019) adopted the concept of transformers [35] to create an attention-augment convolutional block, which required more trainable parameters than that used by [18]. However, these studies focused on enhancing the significant information in convolutional blocks. In this study, we focused on the significant and uncertainty information because the latter might have potential features that can help in improving the model.

## III. Methodology

We designed a novel convolutional attentional block, CUAB. It can leverage the uncertainty information to enhance the convolutional kernel to capture potential information that can improve the model performance. We demonstrate the working of the CUAB and how it can be connected with the existing popular convolutional blocks.

### A. Uncertainty Estimation

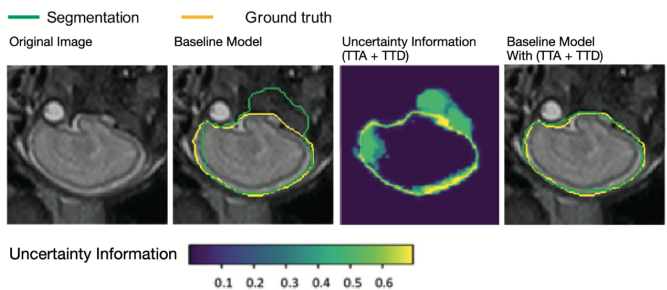

Figure 1. Predicted result of the baseline model (with and w/o TTD and TTA) and the uncertainty information when TTA and TTD are applied. [1].

The concept of CUAB was inspired by TTA and TTD [22], [23]. There are two different types of uncertainty information in a CNN model, namely epistemic and aleatoric. The epistemic information represents the uncertainty of a model and the aleatoric information describes the uncertainty of data, such as noise and randomness in testing data. Thus, TTD can simulate the concept of epistemic and TTA can reimplement the ideas of aleatoric. TTD can leverage the dropout layer during the inference and produce several predictions for each data instance. In contrast, the TTA can retain the model structure and implement data augmentation, such as vertical flipping or rotation, to the testing data. Both methods followed the Monte Carlo simulation to generate several predicted results for the same data; subsequently, the ensemble was adopted to produce the final prediction. The authors of [22] utilized the entropy to quantify the uncertainty information for the testing data, as shown in Figure 1. It was found that a higher value of uncertainty information matches the predicted result with the baseline (TTD + TTA) and closer to the ground truth. In this case, the high uncertainty may have more potential information, which can improve the model performance in a classification or segmentation task. Therefore, we designed the CUAB module by leveraging the uncertainty information to enhance the convolutional block to extract more efficient features.

### B. Convolutional Uncertainty Attention Block

We designed a new convolutional uncertainty attention block by referring to the concept of random forest [36] and Monte Carlo simulation [24], as shown in Figure 2, to add the uncertainty information to the convolutional layer. In the CUAB, we leveraged $P$ $1 \times 1$ convolutional layers with $S$ kernels to produce $X_{P_i} \in \mathbb{R}^{W \times H \times S}, i \in \{1, ..., P\}$. The $P$ denotes the participators that simulated the number of trees in the random forest. We set $S$ to present the number of kernels, which can imagine that is the number of samples in the Monte Carlo simulation. The concept of this structure involves the simulation of a scenario that has $P$ participators, which is similar to the concept of multi-heads in [4], to observe the importance of information from their $S$ feature maps. Next, we adopted global average pooling to $X_{P_i}$ on the spatial-wise (1) to obtain $X_{C_i} \in \mathbb{R}^{W \times H \times 1}, i \in \{1, ..., P\}$, which is the information on the spatial. Here, $X_{C_i}$ represents the certainty information from each participator. We then obtained the uncertainty information $X_{U_i} \in \mathbb{R}^{W \times H \times 1}, i \in \{1, ..., P\}$ using the complement of $X_{C_i}$ in set theory. To consider all uncertainty information from participators $P$, we concatenated all $X_{U_i}$ to obtain $X_U$. Thus far, we focused on the spatial-wise information; however, now, the channel-wise information must be captured. Therefore, we followed the concept of the squeeze-and-excitation network [5] to apply global average pooling to channel-wise (2), and leveraged the $1 \times 1$ convolutional layer with a sigmoid function to obtain $X_{UA} \in \mathbb{R}^{1 \times 1 \times C}$. Finally, we multiplied $X_{UA}$ with $X$ and leveraged the residual path to keep the original input features. The CUAB pseudocode is presented in Table 1.

$$X_{C_m}(i, j) = \frac{1}{S} \sum_{l=0}^{S} X_{P_m}(i, j), \quad \forall m, i, j \in \mathbb{N} \quad (1)$$

$$X_{UA_m} = \frac{1}{H+W} \sum_{i=0}^{H-1} \sum_{j=0}^{W-1} X_{U_m}(i, j), \quad \forall m, i, j \in \mathbb{N} \quad (2)$$

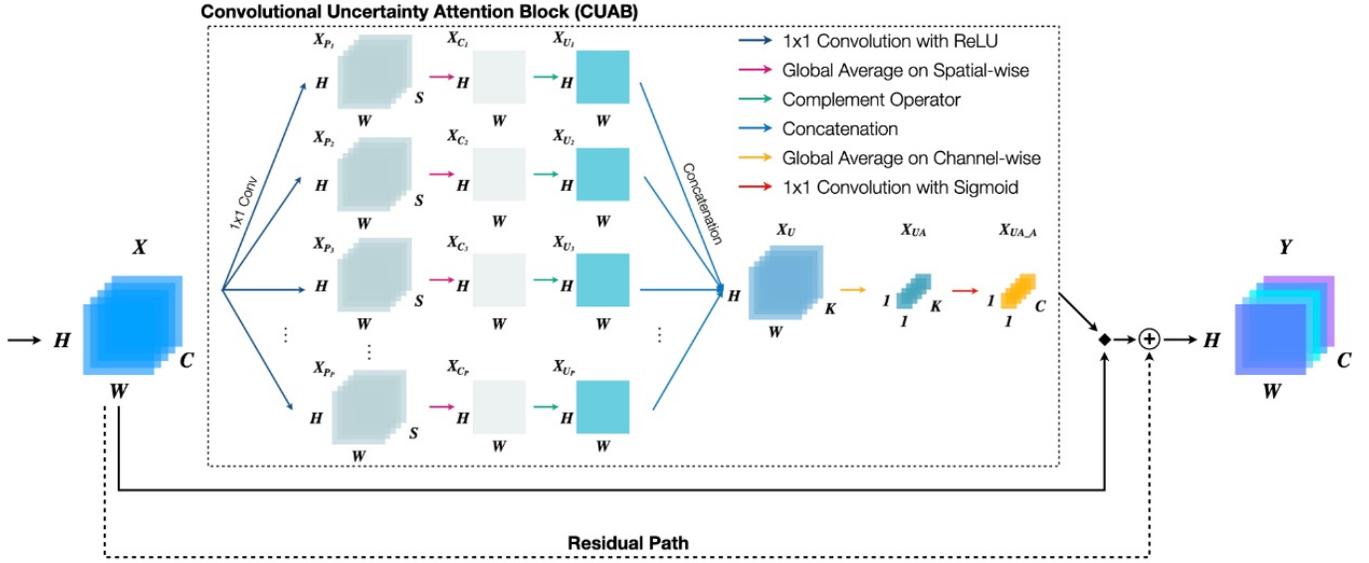

Figure 2. Convolutional Uncertainty Attention Block. We leveraged the multi-head to extract various uncertainty information from the squeezed-and-excitation network to enhance the feature maps

TABLE 1.
PSEUDOCODE FOR THE CONVOLUTIONAL UNCERTAINTY ATTENTION BLOCK.

| | |
|---|---|
| 1 | *# P is the number of participators and S is the number of kernels* |
| 2 | **FUNCTION** Uncertainty_Attention_Foward (X, in_channels) |
| 3 |     *# INPUT: A feature map X and the number of channels in X* |
| 4 |     *# OUTPUT: A new feature Y, which is of the same size as X* |
| 5 |     X_U_TEMP ← EMPTY_LIST [ ] |
| 6 | |
| 7 |     **FOR** i **IN THE RANGE OF** 1 **TO** P: |
| 8 |         X_p ← **The i-th CONVOLUTIONAL**(kernel_size: 1, kernels: S)(X) |
| 9 | |
| 10 |         X_p_active ← **ReLU**(X_p) |
| 11 |         X_c ← **SPATIAL GLOBAL AVG POOLING**(X_p_active) |
| 12 |         X_u ← 1.0 − X_c *# complement in set theory* |
| 13 |         **APPEND** X_u **TO** X_U_TEMP |
| 14 |     **END FOR** |
| 15 | |
| 16 |     X_U ← **CONCATENATE**(X_U_TEMP) |
| 17 |     X_CUAB ← **CHANNEL GLOBAL AVG POOLING**(X_CUAB) |
| 18 | |
| 19 |     X_CUAB_A ← **CONVOLUTIONAL**(kernel_size: 1, kernels: in_channels)(X) |
| 20 | |
| 21 |     X_CUAB_A ← **SIGMOID**(X_CUAB_A) |
| 22 |     Y ← **MULTIPLY** X_CUAB_A **WITH** X |
| 23 | |
| 24 |     **RETURN** Y |
| 25 | **END FUNCTION** |

## C. CUAB in Network

The CUAB is a convolutional attention block that can be easily applied to the existing convolutional blocks (K. He et al., 2016a; Xie et al., 2017), as shown in Figure 3. In Figures 3 (a), we considered improving the input feature map before passing through the residual block. The CUAB directly changes the original input *X* before feeding it to the convolutional block, which is different from Figures 3 (b) and (c). Meanwhile, we put the CUAB to enhance the feature map, which was extracted from the convolutional block in Figures 3 (b) and (c). The only difference between the CUAB combinations in Figures 3 (a) and (b) is that the CUAB is connected before or after the residual path, respectively. These three types of combinations can be used in ResNet (K. He et al. (2016a) and ResNeXt [7]. In contrast, we only adopted the CUAB at the start or end in the convolutional block based on DenseNet [8]. Because it did not have a residual path, it was difficult to implement combination (b) in the dense block. We will show the experimental results and discussion on ResNeXt using Figures 3 (a)–(c) in experiments section.

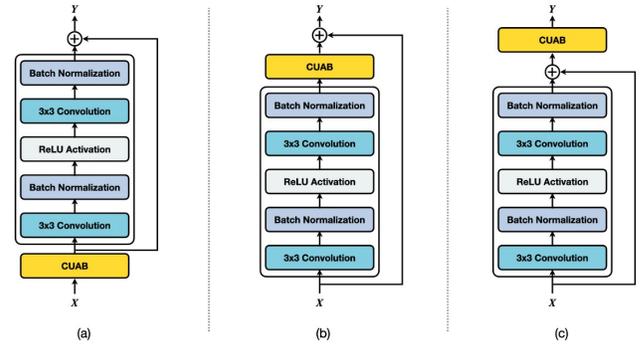

Figure 3. CUAB applied to a residual block as an example. There are three different types of combinations in the residual block: (a) at the start, (b) in the middle, and (c) at the end of the block.

The CUAB aims to capture the potential features from the uncertainty information to enhance the feature maps from the convolutional layer. We utilized the concept of multi-heads to simulate trees from the random forest and adopted an ensemble to obtain the certainty information. Subsequently, we leveraged the complement in the set theory to obtain the uncertainty information. Finally, we followed the concept of the squeeze-and-excitation network to obtain the attention weights for each channel on the original feature maps. The CUAB focused on spatial-wise information and considered the importance of channel-wise features. Moreover, it is a lightweight and flexible convolutional attention block that can easily connect to notable convolutional blocks with fewer additional trainable parameters.

## IV. EXPERIMENTS

### A. Pneumonia Detection by Segmentation Model

TABLE 2.
NCKUH PNEUMONIA DATASET FOR SEGMENTATION AND DETECTION.

| Lung Status | Number of Training Data | Number of Testing Data |
|---|---|---|
| **CXR segmentation data that annotated the location of infected lung.** | | |
| Normal Lung | 578 | 144 |
| Diseased Lung (Pneumonia) | 622 | 151 |
| Total | 1,200 | 295 |
| **CXR detection data that contained the binary classified information.** | | |
| Normal Lung | - | 893 |
| Diseased Lung (Pneumonia) | - | 198 |
| Total | - | 1,091 |

Pneumonia detection is a common but difficult task in chest X-ray (CXR). The most common way for detecting pneumonia involves the creation of a classification model to distinguish between the infected and normal lungs. Then, a class activation map [9] can be leveraged to visualize the location of the infected lung. However, this location produced by the classification is not precise and contains a significant amount of noise. Although the classification model can achieve good results, it cannot provide accurate information regarding the infected locations to aid physicians. Therefore, in this experiment, we designed a pneumonia segmentation model based on U-Net [10] to combine with ResNet-50 and ResNeXt-50 with half channels for each convolutional layer to reduce the computational cost. In the experiment, we reduced the number of channels to half of that of the original ResNet and ResNeXt network. First, we utilized the segmentation model to determine the accurate position of the infected lung and produced a heatmap to present information related to the disease. We then leveraged the information on the heatmap to distinguish between normal and abnormal lungs. In this experiment, the dataset was provided by the National Cheng Kung University Hospital (NCKUH), and the details of the dataset are presented in Table 2. Additionally, we spent a significant amount of time discussing the infected lung annotations with physicians. There are three different characteristics in an infected lung in CXR: consolidation, ground-glass opacity (GGO), and retrocardiac.

Thus, the physicians decided to separate the original annotations into two different classes: consolidation + GGO and retrocardiac.

In the training phase, we utilized data augmentation techniques, such as vertical flipping, rotation, and blur, to increase the data variety. We also applied randomized cropping, zoom in/out, and distortion to increase the difficulty in the training sample. These methods can help in preventing model overfitting and make the model more robust. We used the dice score (5) as the evaluation metric, where $Y \in \mathbb{N}^{N \times W \times H}$ represents the reference masks and $\hat{Y} \in \mathbb{R}^{N \times W \times H}$ denotes the predicted mask. For comparison, we adopted different convolutional attention blocks, such as squeeze-and-excitation (SE), convolutional attention block module (CBAM) [11], and CUAB on the same segmentation model structure. The experimental results are summarized in Table 3.

TABLE 3.
PERFORMANCE OF SEGMENTATION MODEL COMPARISON WITH DIFFERENT CONVOLUTIONAL ATTENTION BLOCKS.

| Model | Total | Pneumonia + GGO | Retrocardiac | Parameters |
|---|---|---|---|---|
| ResNet50 U | 0.6681 | 0.7316 | 0.6046 | 20.19M |
| ResNet50 U CBAM | 0.6103 | 0.7255 | 0.4951 | 21.49M |
| ResNet50 U SE | 0.6759 | 0.7247 | 0.6271 | 20.84M |
| ResNet50 U CUAB | **0.6905** | **0.7439** | **0.6372** | 20.98M |
| ResNeXt50 U | 0.6818 | 0.7280 | 0.6356 | 20.06M |
| ResNeXt50 U CBAM | 0.6520 | 0.7303 | 0.5738 | 21.36M |
| ResNeXt50 U SE | 0.6607 | 0.7381 | 0.5832 | 20.71M |
| ResNeXt50 U CUAB | **0.7305** | **0.7715** | **0.6894** | 20.85M |

$$Dice\ Score(Y, \hat{Y}) = \frac{1}{N}\sum_{i=1}^{N}\frac{2\times|Y_i \cap \hat{Y}_i|}{|Y_i| + |\hat{Y}_i|} \quad (5)$$

After obtaining the predicted infected lung mask, we utilized (6) to distinguish between the normal or abnormal CXR; subsequently, we set the threshold to 0.5. We adopted sensitivity (7) and specificity (8) to evaluate the model. The pneumonia detection performance using the predicted mask is presented in Tables 3 and 4. The receiver operating characteristic (ROC) curve and area under the curve (AUC) of ROC values are presented in Table 4.

$$PenumoniaDetection(\hat{Y}_i) = \begin{cases} 1, if\ \max(\hat{Y}_i) \geq Threshold \\ 0, \quad Otherwise \end{cases} \quad (6)$$

$$Sensitivity = \frac{TP}{TP+FN} \quad (7)$$

$$Specificity = \frac{TN}{TN+FP} \quad (8)$$

<div style="text-align:center">

TABLE 4.
PNEUMONIA DETECTION PERFORMANCE WITH DIFFERENT
CONVOLUTIONAL ATTENTION BLOCKS.

</div>

| Model | Sensitivity (threshold=0.5) | Specificity (threshold=0.5) | AUROC |
|---|---|---|---|
| ResNet-50 | 0.6010 | 0.8589 | 0.8131 |
| ResNet-50 CBAM | 0.7121 | 0.7380 | 0.8105 |
| ResNet-50 SE | 0.7475 | 0.7279 | 0.8028 |
| ResNet-50 CUAB | 0.7677 | 0.7077 | 0.8032 |
| ResNeXt-50 | 0.6212 | 0.8555 | 0.8204 |
| ResNeXt-50 CBAM | 0.7525 | 0.7044 | 0.8019 |
| ResNeXt-50 SE | 0.7020 | 0.7648 | 0.8206 |
| ResNeXt-50 CUAB | 0.7273 | 0.8074 | 0.8439 |

In this experiment, we focused on providing an accurate infected lung mask to the physicians. We then took the predicted mask to create a pneumonia detection model using (6). The results showed that ResNeXt-50 U-Net with the CUAB achieved the highest AUROC score in all models. The results in Tables 3 and 4 demonstrated that the CUAB is an efficient convolutional attention block that only requires a few additional training parameters and improves the model performance in ResNet and ResNeXt.

### B. Pneumothorax Segmentation

To prove that the CUAB can be easily applied to different datasets and improve the model performance, we designed another model to perform pneumothorax segmentation. In this experiment, the model was constructed using ResNet-34 and U-Net. Moreover, we leveraged the feature pyramid network (FPN) [38] and skip-connection integration [39] to improve the long skip connection in U-Net. The pneumothorax dataset provided by NCKUH was annotated by physicians. The details of the dataset are presented in Table 5.

<div style="text-align:center">

TABLE 5.
NCKUH PNEUMOTHORAX DATASET FOR SEGMENTATION.

</div>

| Lung Status | Number of Training Data | Number of Testing Data |
|---|---|---|
| Non-Pneumothorax | 1,998 | 490 |
| Pneumothorax | 1,369 | 354 |
| Total | 3,367 | 844 |

We also adopted data augmentation, similar to pneumonia segmentation and detection. Data augmentation is a useful technique in the training phase if the training data are insufficient. Especially in medical imaging, it is difficult to collect a sufficient amount of data and annotated information. Therefore, data augmentation can efficiently reuse the data and improve the performance and robustness of the model. We adopted the dice score (5) as the metric to evaluate the segmentation model. All models were based on ResNet-34 and U-Net. The only difference between the models is that they utilize a different convolutional attention block to enhance their performance. The results of pneumothorax segmentation are presented in Table 6. Additionally, several real cases for the different convolutional attention blocks in the model are depicted in Figure 4.

<div style="text-align:center">

TABLE 6.
PNEUMOTHORAX SEGMENTATION MODEL COMPARISON WITH DIFFERENT
CONVOLUTIONAL ATTENTION BLOCKS.

</div>

| Model | Dice Score | Parameters |
|---|---|---|
| ResNet-34 U-Net with FPN | 0.82 | 25.4M |
| ResNet-34 U-Net with FPN, CBAM | 0.80 | 26.0M |
| ResNet-34 U-Net with FPN, SE | 0.81 | 25.5M |
| ResNet-34 U-Net with FPN, CUAB | **0.84** | 25.7M |

We found that the convolutional attention blocks SE and CBAM do not work in the segmentation model in Table 6. In contrast, the CUAB apparently improved the performance of pneumothorax segmentation. It also demonstrated a more accurate segmentation mask in Figure 4. In Figure 4 (a), the model with the CUAB can detect more areas on the CXR than the other model with SE and CBAM. The model with the CUAB produced a more complete and accurate predicted mask in comparison to the reference mask in Figures 4 (b)-(c). In this experiment, we leveraged the pneumothorax dataset to evaluate the benefits of the CUAB. The results showed that it can improve the convolutional layer to extract more efficient features to improve model performance. Moreover, we observed that the model with the CUAB only required fewer additional trainable parameters. Thus, the pneumonia and pneumothorax segmentation tasks showed that the CUAB performs well and improves the model performance.

### C. Hyperparameters in CUAB

In this experiment, we discussed two hyperparameters: the participators and samples in the CUAB. The participators presented the number of parallel feature extraction branches in the CUAB. The samples presented the number of convolutional kernels to produce the new feature maps for each participator. Next, we adopted spatial-wise global average pooling to obtain the spatial-wise attention information for the participators. Finally, we concatenated all spatial-wise attention feature maps from the participators. Therefore, the number of participators and samples is an important parameter in the CUAB. We used the pneumonia dataset to evaluate the model with hyperparameters in the segmentation task. We used ResNeXt-50 as the encoder to extract the features from the input images.

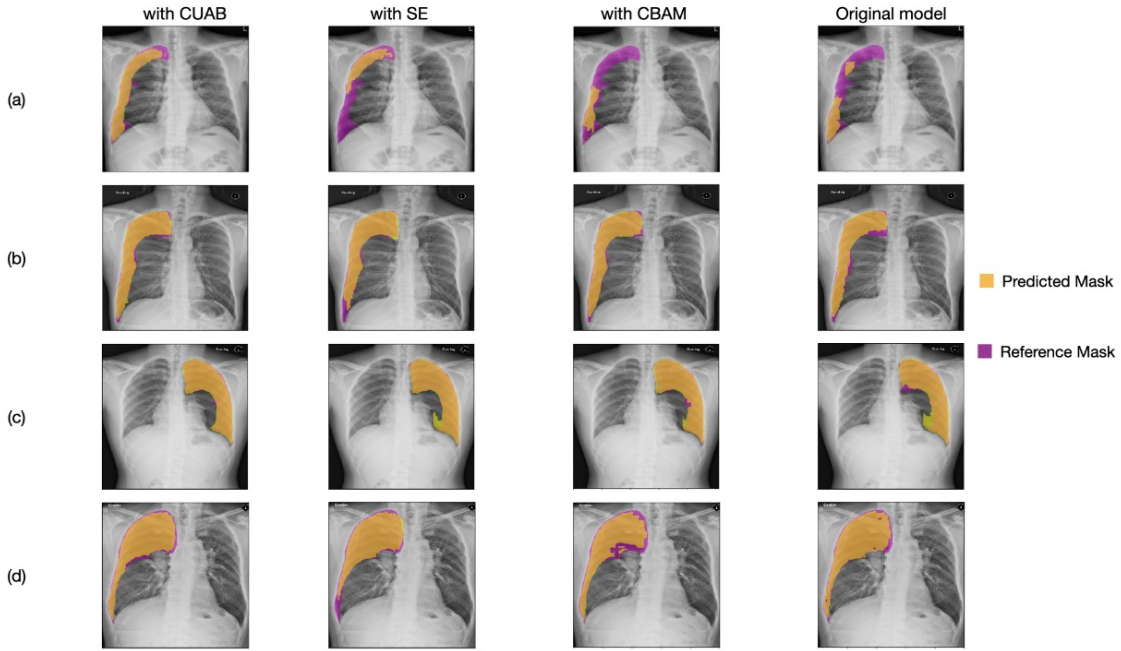

Figure 4. Real cases in the pneumothorax segmentation model with different attentional blocks.

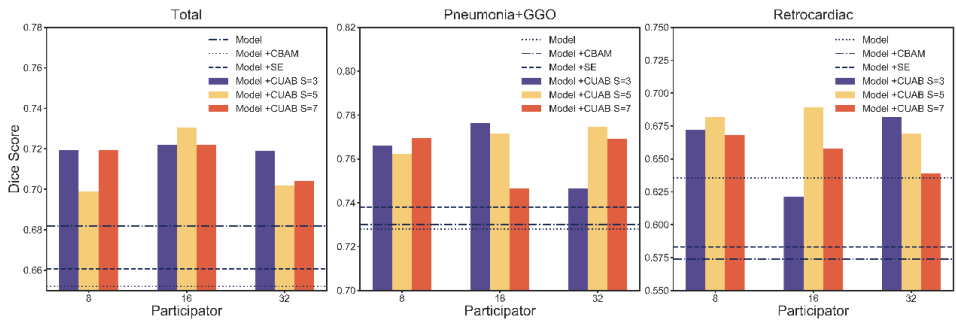

Figure 5. Experimental results on the model with different hyperparameter evaluation in the pneumonia segmentation task.

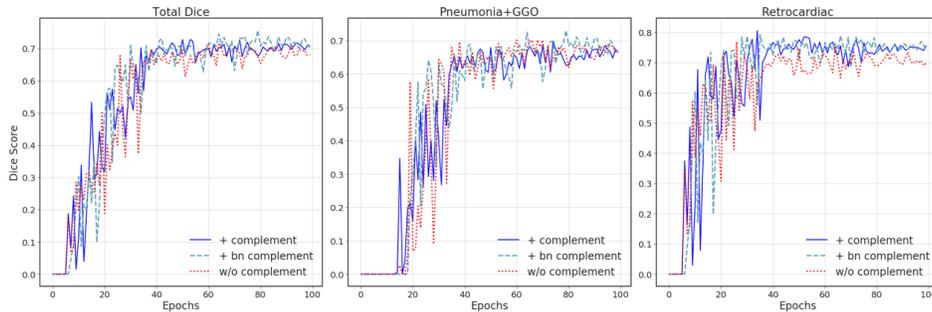

Figure 6. The influence of the complement operator in the CUAB. We presented three different settings on the pneumonia segmentation task in this experiment. The solid blue line is the CUAB with the complement operator and, the blue dot line we added the batch normalized in the CUAB; the solid red line shows the CUAB without the complement operator.

Subsequently, we leveraged U-Net to be the decoder to produce the segmentation mask. In this experiment, $P \in \{8, 16, 32\}$ and $S \in \{3, 5, 7\}$ denote the numbers of participators and samples, respectively. We took different combinations from P and S to evaluate the model and the results are illustrated in Figure 5.

It is evident from Figure 5 that the model with the CUAB outperformed the original model and that with other attention blocks. This is because the structure of the CUAB is similar to the concept of cardinality in ResNeXt [28]. It utilized several convolutional branches to extract various features, and then aggregated that information to improve feature representation. In contrast, we observed that the model performance does not improve as the number of participators or samples increases. The best model performance in the number of participators and samples was observed 16 and 5. However, these parameter settings may change in different tasks. Therefore, we can follow the phenomenon observed from the experimental results to set the number of participators between 8 and 16 with different samples.

### D. Position of CUAB in Network

To consider the position of the CUAB in the existing deep neural network, we designed three different methods to implement the CUAB in the ResNet family, as shown in Figure 3. In this experiment, we discussed three different positions of the CUAB in the model, i.e., Head, Middle, and Tail. We utilized the pneumonia data to evaluate the model performance on the segmentation task. The model encoder was constructed by ResNeXt-50; then, we adopted U-Net as the decoder in the model. The number of participators and samples was set to be 16 and 5 in the CUAB, respectively. The results are presented in Table 7.

TABLE 7.
RESULTS OF THE PNEUMONIA SEGMENTATION MODEL WITH THE CUAB IN DIFFERENT POSITIONS.

| Model | Total Dice | Pneumonia + GGO Dice | Retrocardiac Dice | Parameters |
|---|---|---|---|---|
| ResNeXt50 U | 0.6818 | 0.7280 | 0.6356 | 20.06M |
| ResNeXt50 U CBAM | 0.6520 | 0.7303 | 0.5738 | 21.36M |
| ResNeXt50 U SE | 0.6607 | 0.7381 | 0.5832 | 20.71M |
| ResNeXt50 U CUAB +*Head* | 0.6705 | 0.7407 | 0.6043 | 20.75M |
| ResNeXt50 U CUAB +*Middle* | **0.7305** | **0.7715** | **0.6894** | 20.85M |
| ResNeXt50 U CUAB +*Tail* | 0.7253 | 0.7651 | 0.6855 | 20.85M |

The results showed that the CUAB connected to the convolutional block in the head used less trainable parameters than those located in the middle or tail. However, the model performance was not comparable to that of the CUAB in the middle and tail. This is because the CUAB is an attention block that can enhance the feature maps; if it is placed in the beginning of the convolutional block, it can change the information of the original image. This can result in losing the original information and the model might deteriorate. Therefore, we obtained better model performance when the CUAB was connected in the middle and tail. Both connection methods outperformed the original model and those with other attention blocks.

### E. Complement Process in CUAB

We adopted the complement process to represent the uncertainty information for each branch in the CUAB. Accordingly, we conducted an experiment to observe the influence of the complement process in the CUAB. Figure 6 shows the results of the model with and without the complement in the pneumonia segmentation task. In this task, we divided pneumonia into two different cases based on the position of the symptom; one appears to the position of the lung in the CXR image we called pneumonia+GGO, and the other exists behind the cardiac we named retrocardiac. The experimental results demonstrated that the complement did improve the model performance in this task. In particular, in the retrocardiac class, the CUAB with complement improved the dice score by more than 4% in comparison to the block without the complement. We also adopted batch normalization in the CUAB and obtained a higher performance in this task. Therefore, the complement process is a useful and necessary action in the CUAB.

## V. CONCLUSION

The performance of a CNN model can be improved via three methods, including a deeper and wider network, automatic architecture search, and the attention block module. However, the first two methods require powerful computational devices and it is difficult to design an efficient network block using humans or search strategies and then utilizing this block for deeper and wider stacking. Therefore, in this study, we focused on improving deep neural networks based on attention blocks. Previous studies utilized attention blocks to enhance the significant information on the feature maps. We observed that TTA and TTD utilized the uncertainty information to improve the model performance in the inference phase. Therefore, we designed a novel CUAB to enhance the significant features and capture the potential information in uncertain areas. The experimental results showed that this CUAB outperformed other state-of-the-art attention blocks in different tasks. The CUAB is a lightweight and flexible block that can be easily applied to other popular neural networks, such as ResNet, ResNeXt, and DensseNet. Although the CUAB model achieved higher performance, it still has room for improvement. To consider additional trainable parameters, we applied global average pooling to squeeze the spatial-wise attention information. This wastes the spatial information extracted by the participators. Thus, we considered separating the spatial-wise and channel-wise information into different branches, and then used the concept of highway networks [10]. Additionally, concatenation is another high computational operator that can be improved in the future. We can utilize the idea of self-attention [35] instead of concatenation. This method provided more information to visualize the operations in an attention block.

## ACKNOWLEDGMENT

The authors have no conflicts of interest to declare. This research work was supported by the Ministry and Technology, Taiwan, R.O.C. under Grant MOST 109-2745-B-006-002-, and the authors thank the National Center for High-Performance Computing, Taiwan for providing computational and storage resources. The authors express their appreciation for excellent technical assistance in CUAB performance evaluations from Kun-Chi Kuo and Li-Jin Huang.